\documentclass[conference]{IEEEtran}
\IEEEoverridecommandlockouts
\usepackage{cite}
\usepackage{amsmath,amssymb,amsfonts}
\usepackage{algorithm}
\usepackage{algpseudocode}
\usepackage{graphicx}
\usepackage{booktabs}
\usepackage{array}
\usepackage{xcolor}
\usepackage{listings}
\usepackage{scratch3}
\usepackage{url}
\usepackage[hidelinks]{hyperref}
\usepackage{microtype}

\newcommand{\sys}{SchedCheck}

\newcommand{\sched}{\ensuremath{\mathit{Sched}}}
\newcommand{\lensto}{\ensuremath{\sqsubseteq}}

\newtheorem{theorem}{Theorem}
\newtheorem{definition}{Definition}
\newtheorem{lemma}{Lemma}

\begin{document}

\title{\sys{}: Schedule-Robustness Analysis for\\Event-Driven Block Programs}

\author{
\IEEEauthorblockN{Yuan Si and Jialu Zhang$^{*}$\thanks{Corresponding author: Jialu Zhang.}}\\
\IEEEauthorblockA{University of Waterloo, Waterloo, Canada\\
\texttt{yuan.si@uwaterloo.ca}, \texttt{jialu.zhang@uwaterloo.ca}}
}

\maketitle

\begin{abstract}
Block-based languages such as Scratch let beginners assemble interactive programs from sprites and scripts. These programs are concurrent in practice: green-flag scripts, broadcasts, and clones run as cooperatively scheduled threads over shared sprite and stage state, and their authors never write a thread. We show that such programs contain schedule-sensitive behaviors whose observable result depends on an execution order the language leaves open. Editing, saving, or remixing a project can produce a copy with the same blocks but a different layer order, changing the order the virtual machine starts scripts. We formalize the schedule space a Scratch virtual machine can realize as the permutations of the initial executable-target order, and define schedule-robustness against a lattice of observation lenses over a fixed horizon. A partial-order exploration runs one schedule per dependence-equivalence class, and on projects small enough to enumerate, an independent oracle confirms it recovers every realizable outcome. On larger projects, representatives stand in for the factorial under the validated dependence model. \sys{} implements this on the production Scratch VM. Across 224 real student projects, at least 21\% of the concurrent ones are schedule-sensitive at the grading lens, and a uniform random sample of public projects replicates the rate at 17.6\%, with two real remixes of a deployed animation arranging its letters differently. On hand-built fault pairs and a generated benchmark of 32 spec-defined faults across four classes, the tool detects and localizes every schedule fault, with a logic-fault control reporting clean. The oracle exposed four unsoundness gaps in the dependence model, all repaired. The method is parametric in the execution model, instantiating unchanged on a second cooperative event loop.
\end{abstract}

\section{Introduction}
\label{sec:intro}

Scratch is the entry point to programming for tens of millions of children. A learner drags blocks onto sprites, clicks the green flag, and watches a scene come alive. Underneath the friendly surface the runtime is concurrent. Every sprite can carry several scripts that start together on the green flag, react to broadcasts, or run inside clones, and all of them read and write shared state: variables, lists, and the position, costume, and visibility of every sprite. The learner writes interacting threads without ever naming one.

Consider a child who builds a small game where a ball drops onto a paddle and a counter records the catch: the paddle checks whether it touches the ball and sets a variable, and the ball moves to the paddle. The child runs it, sees the catch register, and shares it. A classmate remixes the project, opening it and saving a copy, which carries the two sprites in a different layer order, and now the paddle sometimes checks for contact before the ball has moved. Same blocks, same inputs, same random seed, yet the counter reads zero, and nothing in the project is a recognizable bug. The behavior depends on the order in which the two green-flag scripts start, and that order is not fixed by the language.

This is a schedule-sensitive fault, and the schedule it depends on is realizable by a routine user action. A Scratch virtual machine starts green-flag scripts in a deterministic sweep over the sprites in layer order, so the relative order of two sprites' scripts is exactly the relative order of the sprites on the stage. Reordering sprites is a one-click operation in the editor, and saved copies of a project can differ in it. A program whose result changes under such a reordering can carry a defect that single runs, fixed inputs, and code review all miss.

Existing analyses for Scratch look elsewhere. Test generators such as Whisker~\cite{stahlbauer2019whisker} drive a project with synthesized inputs and check acceptance under one execution per input; static checkers such as LitterBox~\cite{fraser2021litterbox} flag bug patterns and code smells in the block graph. Both target correctness in the space of inputs or the shape of the code, and neither perturbs the schedule. Classical concurrency testing does~\cite{flanagan2005dpor,abdulla2014optimal}, yet it explores interleavings of a fixed or dynamically spawned thread set, a model that overshoots the single ordering choice a Scratch VM actually exposes.

We take the schedule itself as the object of testing. The realizable schedule space of a Scratch program under a fixed input and seed is the set of runs obtained by permuting the initial inter-sprite order, a finite space the production VM can reproduce through a layer reordering. We define schedule-robustness of a program against an observation lens as agreement of the observed behavior across this space. The lens hierarchy, from the final state up to the full per-tick trace, is the one existing work uses to compare two Scratch programs~\cite{scratchlens}; we reuse it to ask whether one program agrees with itself across its schedules, so a verdict states what an order change can and cannot affect. A program is schedule-sensitive at a lens when two realizable orders disagree there, and the witness is two concrete runs.

Exploring the space naively costs a factorial number of executions, and two algorithms make the exploration sound and legible. The first is a partial-order reduction over a typed dependence relation: orders that agree on every dependent pair are equivalent, the equivalence classes are the acyclic orientations of the conflict graph, and the explorer runs the VM once per class, not $k!$ times. The second explains a difference, keeping the observables an order moves, subtracting the order sensitivity a paired reference carries, and scoping what remains to the two scripts that contend for a resource. An independent oracle enumerates all orders on small projects and confirms the reduction recovers the same outcomes, a check that exposed four dependence gaps, two on the course corpus, one on the public sample, and one the broadcast and clone reduction surfaced, all of which we repaired.

\sys{} realizes this design on the production Scratch VM. We evaluate it on 224 real student projects and a labeled benchmark, and we report seven findings. Schedule-sensitivity is common: at least 21\% of the concurrent projects change their final-state observable under some realizable order, and 25\% change a finer observable; a uniform random sample of public projects replicates this at 17.6\%. The reduced exploration is sound in practice, matching exhaustive enumeration with no missed outcome on all 142 course and 77 public projects an oracle could check, where it caught and we repaired four footprint gaps. On three hand-built admissible fault pairs the tool detects and localizes all three, and stays silent on a live logic fault that perturbs no schedule. Schedule perturbation reaches faults that input generation and static smell checkers leave untouched. A repaired program re-certifies as schedule-robust while preserving its intended behavior. The method is parametric in the execution model, its model-level parts instantiating unchanged on a second cooperative event loop. The orders it flags do occur between real saved copies: the remix family of a deployed logo animation assembles its letters in different places under two orders its copies save. Such manifestation is rare, though, and most sensitive projects stay latently fragile.

This paper makes the following contributions. It formalizes the realizable schedule space and schedule-robustness of an event-driven block runtime, and proves that every schedule the explorer reports is reproducible on the unmodified VM. It backs the exploration with an independent oracle: on every project small enough to enumerate, the oracle runs all orders and confirms the reduction misses no outcome, so the verdict there is checked exhaustively, and the partial-order reduction carries the same model-based verdict to projects too large to enumerate, one representative per dependence class. It gives a set-valued, site-scoped attribution that reduces a schedule-sensitive difference to the pair of scripts and the resource behind it, avoiding the false negatives of a boolean divergence check. It implements \sys{} on the production Scratch VM, quantifies schedule-sensitivity across a course corpus and a random sample of public projects, exhibits the perturbed order between real saved copies and measures how seldom remixing realizes it, measures detection and localization against seeded faults and a static-analysis baseline, and reports the implementation, benchmark design, and evaluation results.

\section{Background}
\label{sec:background}

\subsection{The Scratch execution model}

A Scratch project~\cite{maloney2010scratch,resnick2009scratch} is a set of targets, one stage and a number of sprites. Each target owns scripts, and each script begins with a hat block that names the event that starts it: the green flag, the receipt of a broadcast, or the creation of a clone. A target also owns mutable state that other scripts can observe: local and global variables and lists, and the sprite's position, direction, costume, size, visibility, and graphic effects. Scripts that run at the same time share this state, which makes a project a concurrent program over shared memory even though its author manipulates only blocks.

The runtime advances in ticks at thirty frames per second. Within a tick the virtual machine steps each active script until the script yields, and a script yields at a loop boundary, a timed wait, the join of a broadcast-and-wait, or a screen refresh. A forever loop runs one iteration per tick and lets every other script run in between. Execution is cooperative: a script holds the machine until it chooses to yield, and the machine never preempts it mid-step.

Concurrency enters through three event sources. The green flag starts every green-flag script at once. A broadcast starts the scripts that listen for its message, and a broadcast-and-wait additionally holds the sender until those scripts finish. A clone block copies a sprite and starts the copy's clone-start scripts. The three sources differ in when they create threads, yet they share the same shared state, so two scripts started by different events can race on a variable or a sprite property exactly as two green-flag scripts can. The green flag is the common case and the one a remix perturbs, and it anchors the model that follows; broadcasts and clones extend the same shared-state picture.

\begin{figure}[t]
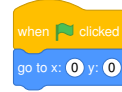

\centering
\begin{minipage}[t]{0.5\linewidth}
\centering
\textsf{Paddle (Sensor)}\par\smallskip
\begin{scratch}[scale=0.5,else word=else]
\blockinit{when \greenflag clicked}
\blockifelse{\boolsensing{touching \selectmenu{Ball}?}}
{\blockvariable{set \selectmenu{caught} to yes}}
{\blockvariable{set \selectmenu{caught} to no}}
\end{scratch}
\end{minipage}\hfill
\begin{minipage}[t]{0.48\linewidth}
\centering
\textsf{Ball (Mover)}\par\smallskip
\begin{scratch}[scale=0.5,else word=else]
\blockinit{when \greenflag clicked}
\blockmove{go to x: \ovalnum{0} y: \ovalnum{0}}
\end{scratch}
\end{minipage}
\caption{The ball-and-paddle project. Under $(\textsf{Ball},\textsf{Paddle})$ the ball moves before the test and \textsf{caught} becomes \textsf{yes}; under $(\textsf{Paddle},\textsf{Ball})$ the test runs first and it becomes \textsf{no}. The orders differ only by sprite layout, which a remix changes.}
\label{fig:example}
\end{figure}

Figure~\ref{fig:example} shows the two scripts of the ball-and-paddle project. Reading them together makes the dependence visible: the paddle's touch test reads the ball's position, and the ball's move writes it, so the two scripts contend for the ball's pose and their order decides the test.

\subsection{The scheduler and its single degree of freedom}

The production VM keeps active scripts in an ordered list and, on every sub-tick pass, steps them in list order. The order of this list at the start of a run is the only place where a choice enters. When the green flag fires, the VM sweeps the targets in layer order and appends each target's green-flag scripts to the list in block-storage order. The relative order of two sprites' scripts in the list equals the relative order of those sprites on the stage. Everything after this point is a deterministic function of the list and the program. Clones and freshly matched broadcast receivers append to the end of the list as they appear, and a restarted receiver keeps its position, so no later event reopens the ordering choice.

This pins down what a perturbation may change. Layer order is editable: a user sends a sprite forward or backward, or remixes a project that was saved with a different layout, and the green-flag sweep then produces a different starting list. The within-target order of several scripts on one sprite is fixed by block storage and is not separately editable. The realizable freedom is the permutation of the initial inter-sprite order, and a sprite that the stage always renders at the back contributes its scripts as a fixed suffix.

We hold two further sources of variation fixed throughout. The pseudo-random generator is seeded, so a draw depends only on the number of prior draws. The clock advances by a fixed amount per tick, so a reader of the timer sees a value tied to the tick and not to how much work ran. These choices isolate the schedule as the variable under study, and they supply the determinism a tester needs to attribute a difference to the order alone.

\section{Schedules and Robustness}
\label{sec:model}

This section fixes what a schedule is, which schedules a run can realize, and what it means for a program to be robust against them. We write $P$ for a project, $\iota$ for a fixed input and IO state, and $r$ for a fixed seed.

\subsection{The realizable schedule space}

A configuration is the ordered executable-target list, the per-target thread lists, the shared state, and the set of pending events. The shared state holds the variables and lists and, for every sprite, its position, direction, costume, size, visibility, and effects. A step runs one thread to its next yield, which reads and writes a bounded set of resources and may append a thread for a clone or a receiver. A tick is a sequence of steps in list order until the threads quiesce, after which the clock advances and the observation is recorded. With the seed and clock pinned, a tick is a deterministic function of the configuration, so a run over a fixed horizon of $H$ ticks is a deterministic function of the program, the input, the seed, and the initial executable-target order.

The one free choice is the initial order of the executable targets, the layer order the VM sweeps when it starts scripts; Section~\ref{sec:background} fixes everything else. Let $\pi$ range over the permutations of the $k$ contributing sprite targets, the executable sprite targets with a hat that can fire under the fixed input or is reachable from one through modeled broadcast and clone edges, the stage held fixed; including an inert target only adds equivalent orders. The order $\pi$ determines the green-flag thread list, and equally the later broadcast and input fan-outs, which the VM appends in the same executable-target order; a clone's scripts start at a position its creator fixes, so two clones spawned in one tick run in their creators' layer order, which makes clone scheduling a function of $\pi$ and not a choice beyond it. Writing $\mathrm{exec}_H(P,\iota,r,\pi)$ for the $H$-tick run under initial order $\pi$, the realizable schedule space is
\[
  \sched(P,\iota,r,H) \;=\; \{\, \mathrm{exec}_H(P,\iota,r,\pi) \;:\; \pi \in S_k \,\}.
\]
The space is finite, and every element is a genuine run of the unmodified VM: assigning the sprites the layer order $\pi$ produces exactly this run.

\begin{theorem}[Realizability]
\label{thm:realizable}
For every $\pi \in S_k$ there is a layer assignment under which the unmodified VM, on the same $P,\iota,r$, produces $\mathrm{exec}_H(P,\iota,r,\pi)$. Consequently any behavioral difference \sys{} reports between two explored schedules is exhibited by two real runs that differ only in sprite layer order.
\end{theorem}

The argument rests on how the VM starts scripts. On a green-flag, broadcast, or input event, the VM matches hats by sweeping the executable targets in layer order~\cite{scratchvm} and appends the matching threads in that order, so the relative order of any two targets' threads, at the first fan-out and at every later one, is the relative layer order of those targets. Assigning the sprites the layer order $\pi$ therefore reproduces $\pi$ at every such fan-out at once. Clone creation starts the new clone's scripts at a position fixed by its creator, so two clones spawned in one tick run in the layer order of their creators, and clone scheduling is a function of $\pi$ rather than a choice beyond it; a broadcast-and-wait join records the receivers' identities, not their list positions, so the reordering preserves it; and in-place restart is a function of the run so far. Hence $\mathrm{exec}_H(P,\iota,r,\pi)$ is a run of the unmodified VM on the layer-reordered project, by the deterministic sweep of Section~\ref{sec:background}.

Two consequences shape the rest of the paper. The space is the symmetric group $S_k$, not an unbounded tree of interleavings, which is what makes a complete exploration tractable. The stage is a fixed suffix, because it is rendered at the back and no layer change moves it past a sprite, so permuting it would describe a run the VM cannot produce.

\subsection{Observation lenses}

What counts as a difference depends on what an observer watches. We reuse the observation lenses of existing work~\cite{scratchlens}, four that form a chain from coarse to fine. The final-state lens reports the variables, lists, and sprite poses at the horizon. The frame-visible lens adds the per-tick rendered poses. The monitors lens adds the on-stage variable and list watchers. The full-trace lens adds the clone and event history. Each coarser observation is a projection of the next finer one, written
\[
  \text{final} \;\lensto\; \text{frame} \;\lensto\; \text{monitors} \;\lensto\; \text{full},
\]
so agreement at a finer lens forces agreement at every coarser one. The lens is part of a verdict because it states the strength of the claim: a difference at the full-trace lens may be invisible to a player, while a difference at the final-state lens changes the result the project is built to produce.

\begin{lemma}[Monotonicity]
\label{lem:mono}
If two runs agree under a finer lens, they agree under every coarser lens. A program robust at a finer lens is robust at every coarser lens.
\end{lemma}

The lemma follows from the projection structure: each coarser observation is a function of the next finer one, so equality of the finer observations carries to the coarser. The practical reading is a decay curve. As the lens coarsens from the full trace to the final state, the set of schedule-sensitive projects can only shrink, and the gap between the two ends measures fragility that an order change introduces during a run yet that the program resolves before it ends.

\subsection{Schedule-robustness}

\begin{definition}[Schedule-robustness]
$P$ is schedule-robust at lens $L$ for $(\iota,r)$ over horizon $H$ when all schedules in $\sched(P,\iota,r,H)$ agree under $L$. $P$ is schedule-sensitive at $L$ when two of them disagree.
\end{definition}

Robustness is monotone along the chain by Lemma~\ref{lem:mono}, so a verdict targets the grading lens, the lens at which a project's intended result lives, and reports the coarser robustness for free. The ball-and-paddle project of Section~\ref{sec:intro} is schedule-sensitive at the final-state lens, since the counter variable takes two values across the two sprite orders. A witness is the pair of orders together with the values they produce, and the witness runs on the stock VM.

Schedule-robustness restates a classical idea for this setting. A program robust at the full-trace lens is one whose realizable schedules all produce the same run up to the reordering of independent work, which is the determinacy that Bernstein's conditions characterize for two pieces of work~\cite{bernstein1966} and that Mazurkiewicz trace theory lifts to a set of events~\cite{mazurkiewicz1987trace}. The lens lattice grades this notion: a program can be determinate at the level a grader observes while two of its schedules still differ in an order an instrument could record. A verdict carries both a yes-or-no answer and the lens at which it holds, and the next sections turn the definition into a procedure that decides it and a reduction that decides it cheaply.

\section{Testing for Schedule Sensitivity}
\label{sec:testing}

\sys{} tests a project by running it under several realizable schedules and comparing what it observes. The loop fixes the input and seed, chooses an initial sprite order, sets the green-flag thread list to that order, and runs the VM to a fixed horizon. It records the observation at each lens. A project is schedule-sensitive at a lens when two orders produce different observations there, and the two orders form the witness.

\subsection{A determinism harness}

A reported difference is only meaningful when the order is the one thing that changed. The harness removes every other source of variation. It seeds the random generator so a draw is a function of the number of prior draws, and it gives identifier allocation its own seeded counter so clone and variable identities do not drift between runs and shift the values that depend on them. It freezes the clock to advance by one tick of simulated time per step and routes the sequencer's work budget through this clock, so the number of script steps in a tick follows from the program and not from wall-clock time. It pins the project timer and the calendar reads that some sprites consult, which otherwise vary between runs. Advancing the clock per tick, not per read, keeps it from coupling to the interleaving while still letting timer-driven behavior progress.

The harness guards every measurement with a determinism check. It runs the identity order twice and requires the two full traces to agree exactly. A project that fails this check has a residual source of nondeterminism that the harness does not pin, and \sys{} marks it untrusted and leaves its differences unattributed. Across the corpus this check fails on no project, which gives confidence that a difference under reordering is a scheduling effect.

\subsection{Witnesses}

A witness is two sprite orders and the observations they produce, and both run on the unmodified VM by Theorem~\ref{thm:realizable}. A bare witness names two outcomes without explaining them. \sys{} turns it into a cause: the conflict graph names the pair of scripts whose order the witness flips and the resource they contend for, a write and a read of the same variable, a broadcast whose receiver the sender no longer waits for, or a sprite that reads another's pose before that sprite has moved. On a labeled control-mutant pair the set-valued attribution of Section~\ref{sec:dpor} sharpens this to the exact observables the fault makes order-sensitive. The contended pair is the information a fix has to act on and the diagnosis a learner can read.

\subsection{Fault classes}

Schedule-sensitive faults in Scratch recur in four shapes, and the names describe the contended resource. A missing broadcast-and-wait lets a sender continue before the receiver it triggered has finished, so a value the receiver was to prepare is read early; the contended resource is the variable the receiver writes. A read before initialization has one green-flag script read a variable that a second green-flag script initializes, and the order of the two scripts decides whether the read sees the initial value. A clone-initialization race uses a clone before the clone's own setup has run, so the clone is observed in a half-built state. A sensing race has a sprite test another sprite's position or contact before that sprite has moved, which is the shape of the ball-and-paddle project. The four classes share a structure, a write and a read of one resource whose order the language leaves open, and they organize the benchmark and the witnesses the tool reports.

\subsection{Admissible faults for measurement}

To measure detection we need faults whose ground truth we control. A seeded fault pairs a control program with a mutant. The pair is an admissible schedule fault when three conditions hold. The mutant's effect must be observable at some lens. The mutant must agree with the control under the identity order, so a fault that changes behavior under every order is a logic fault, not a schedule fault, and falls outside this measurement. The disagreement must appear under some realizable initial order, which excludes a fault that only a sub-tick interleaving could expose. The control plays the role of the repaired program, and the gap between control and mutant under reordering is the fault we want a tool to find.

\section{Exploring and Explaining the Schedule Space}
\label{sec:dpor}

The schedule space is finite, yet enumerating $S_k$ costs $k!$ runs and most of
those runs agree. This section gives the two algorithms at the center of the
approach. The first explores the space at one run per dependence-equivalence
class, not $k!$ runs, and recovers every distinct observable
outcome. The second takes a difference the first reports and reduces it to a
cause, the pair of scripts whose order decides the outcome and the observable they
fight over. The two together turn a factorial search into a small, complete
exploration that ends in a diagnosis.

\subsection{Footprints and dependence}

Each sprite has a footprint, a typed record of the resources its scripts
read, write, create, delete, and consume: variables and lists, the sprite properties
that affect rendering and sensing, clone families, broadcast channels, and the ordered
tokens of the random stream and the timer. The typing is what makes the relation
precise: a write to one sprite's position does
not conflict with a read of another sprite's score, and the explorer keeps the two
sprites apart, while a touch test against a moving sprite reads that sprite's
position and the two stay together. A footprint
over-approximates by design, in the manner of a sound abstract
interpretation~\cite{cousot1977abstract}: when a block's effect is uncertain, the
footprint records the extra accesses, so the relation errs toward dependence and the
reduction stays sound.

In the ball-and-paddle project the paddle's footprint reads the ball's position and
the ball's footprint writes it, so the two conflict on the ball's pose and the
explorer keeps both orders; a third sprite that only animated a backdrop would
conflict with neither and drop out.

Two sprites are independent when their footprints do not conflict. A conflict is a
write against a read or write of the same resource. Two writes of the same variable
through a commutative operation, such as two scripts that each change a score by
one, are exempt from the write-write case, since their order leaves the final value
unchanged up to the observation tolerance, which absorbs the low-order differences
that floating-point summation introduces. The exemption is local: a script that both changes and reads the same
variable is order-sensitive, because its read observes a different value depending
on when the other write lands, so a read of a commutatively written resource
conflicts. An opcode that the table cannot model soundly is given a footprint that
conflicts with everything, which keeps the relation an over-approximation.

\subsection{The reduced exploration}

Swapping two adjacent independent sprites leaves the observation unchanged, which is
the partial-order analog of commuting two independent events. The equivalence this
generates on $S_k$ has a clean description: two orders are equivalent exactly when
they orient every conflict edge the same way, that is, when for every dependent pair
of sprites they agree on which sprite comes first.

\begin{lemma}[Class signature]
\label{lem:signature}
Let the signature of an order be the orientation it gives to each conflict edge.
Two orders are equivalent under independent adjacent swaps if and only if they have
the same signature. The number of classes equals the number of acyclic orientations
of the conflict graph.
\end{lemma}

The forward direction holds because an independent adjacent swap exchanges two
sprites that share no conflict edge, so no edge changes orientation and the
signature is unchanged. The reverse direction sorts one order into the other. Take
two orders with the same signature and bring the first toward the second by
selection: at each position, move the sprite that the second order places there
leftward by adjacent swaps. A swap that the sort needs exchanges two sprites that
the second order has not yet fixed, and if those two were dependent their conflict
edge would force the same orientation in both orders, which the sort would already
respect, so the swap never crosses a dependent pair. Every swap is therefore
independent, and the two orders are connected by independent swaps. Conversely every
acyclic orientation is realized by some order: a topological sort of its directed
edges, breaking ties by sprite index, is a permutation whose signature is that
orientation. The signature is thus a bijection between classes and acyclic
orientations, which turns the number of classes into a combinatorial quantity, the
number of acyclic orientations of the conflict graph, and tells the explorer what to
enumerate.

Algorithm~\ref{alg:explore} enumerates one order per class without ever materializing
$S_k$. The lemma identifies a class with an acyclic orientation of the conflict
edges, so the explorer walks those orientations directly. It fixes the conflict
edges in an order and decides them one at a time; at each edge it tries both
directions and keeps a direction only when the partial orientation so far stays
acyclic, which prunes the cyclic combinations that no permutation realizes. When all
edges are oriented, the canonical linear extension of the resulting partial order,
the lexicographically least order consistent with it, is the representative, and the
VM runs once on it. Independent pairs never appear among the edges, so their order
is left to the canonical extension and never multiplies the work.

\begin{algorithm}[tb]
\caption{\textsc{Explore} runs the VM once per equivalence class by walking the
acyclic orientations of the conflict graph.}
\label{alg:explore}
\small
\begin{algorithmic}[1]
\Require project $P$, input $\iota$, seed $r$, grading lens $L$
\Ensure  $\mathit{obs}$: one witnessing order per distinct outcome
\State $g_1,\dots,g_k \gets$ contributing sprite targets, stage fixed
\State $E \gets \{\,(i,j) : i<j,\ \textsc{Conflict}_L(\mathrm{fp}(g_i),\mathrm{fp}(g_j))\,\}$
\State $\mathit{obs} \gets$ empty map
\State \Call{Walk}{$1,\ \varnothing$}
\State \Return $\mathit{obs}$
\Statex
\Procedure{Walk}{$t,\ \omega$}
  \If{$t > |E|$}
    \State $\pi \gets \textsc{LexMinExtension}(\omega)$
    \State $o \gets$ observe $\mathrm{exec}(P,\iota,r,T_0(\pi))$ at lens $L$
    \If{$o \notin \mathit{obs}$}
      \State $\mathit{obs}[o] \gets \pi$
    \EndIf
    \State \Return
  \EndIf
  \State $(i,j) \gets E[t]$
  \For{$d \in \{\,i \to j,\ \ j \to i\,\}$}
    \If{$\omega \cup \{d\}$ is acyclic}
      \State \Call{Walk}{$t+1,\ \omega \cup \{d\}$}
    \EndIf
  \EndFor
\EndProcedure
\end{algorithmic}
\end{algorithm}

The cost is what the design is for. The walk visits one leaf per acyclic
orientation, and each cyclic prefix is cut the moment it closes a cycle, so the
number of leaves equals the number of classes and the enumeration is
output-sensitive: it runs the VM once per class and not $k!$ times. The
acyclicity test and the canonical extension are a pass over the edges, so the VM
runs dominate the cost. A fully
dependent set is the worst case, where every order is its own class and the walk
reduces nothing; a project whose conflict graph is that dense is capped to a sample,
the only place the search is not exhaustive.

\begin{theorem}[Soundness and completeness]
\label{thm:dpor}
Under a footprint that over-approximates the real accesses, with independence taken
at the grading lens $L$, every order in a class produces the same observation at $L$
as the class representative, and the explorer reports an observation for every order
in $\sched(P,\iota,r,H)$. No realizable outcome is missed, and no reported outcome is
spurious.
\end{theorem}

Soundness comes from the commutation of independent work, with independence the
lens-indexed relation $\textsc{Conflict}_L$ the explorer computes. Two adjacent
sprites it calls independent commute for one of two reasons: their transactions touch
disjoint resources, the Bernstein condition~\cite{bernstein1966} under which neither
reads what the other writes, or they share only commutatively updated resources whose
combined value is order-independent. The second case is value commutativity, not
Bernstein independence, so $\textsc{Conflict}_L$ admits it only up to the monitor
lens, which reads values not operation order, and withholds it at the full-trace lens.
Either way the swap preserves the observation at $L$, and a block move decomposes into
such swaps, so the whole reordering preserves it, by induction over the ticks.
Completeness comes from Lemma~\ref{lem:signature}. The walk
emits one representative per acyclic orientation, every order shares its signature
with exactly one orientation the walk emits, and the dynamic creation of clones and
receivers adds no ordering freedom because their positions follow from the initial
order. The finite ordering choice replaces the unbounded interleaving tree that a
general concurrency tester reasons about, which is what lets a single representative
settle an entire class.

The oracle confirms the lens-bounded exemption loses nothing: the reduced and
exhaustive observation sets agree at every lens, including the full-trace lens where
the exemption is withheld. The footprint conditions the theorem assumes, over
dynamic fan-out, the random and timer streams, unmodeled opcodes, and value-commutative
writes, are the soundness contract of Section~\ref{sec:impl}.

\subsection{Attributing a difference to its cause}

A bare difference names two outcomes without explaining them. Algorithm~\ref{alg:attribute}
turns the difference into an attribution, a set of observable keys that the order
makes the program disagree on, scoped to the pair of scripts responsible.
The explorer already records, for each project, which observable keys move under a
family of realizable orders. A key names what changed and where: a variable by
scope and name, a sprite pose by field, a monitor by identity, a clone by its step
in the population trace. The algorithm computes this diverging-key set for the
suspect program and for a paired reference, then subtracts.

\begin{algorithm}[tb]
\caption{\textsc{Attribute} reduces a difference to its cause by subtracting the
order sensitivity the paired reference already carries.}
\label{alg:attribute}
\small
\begin{algorithmic}[1]
\Require control $C$, suspect $M$, contended-site descriptor $s$, lens $L$
\Ensure  the observable keys the fault makes order-sensitive, scoped to $s$
\State $\Sigma \gets \{\,\mathrm{identity},\ \mathrm{reverse},\ \mathrm{shuffle}_1,\dots\,\}$
\For{$X \in \{C,\ M\}$}
  \State $b \gets$ keyed observation of $\mathrm{exec}(X,\mathrm{identity})$ at lens $L$
  \State $D_X \gets \emptyset$
  \For{$\sigma \in \Sigma \setminus \{\mathrm{identity}\}$}
    \State $o \gets$ keyed observation of $\mathrm{exec}(X,\sigma)$ at lens $L$
    \State $D_X \gets D_X \cup \{\,\kappa : o[\kappa] \neq b[\kappa]\,\}$
  \EndFor
\EndFor
\State $A \gets D_M \setminus D_C$
\State $A_s \gets \{\,\kappa \in A : \textsc{Downstream}(\kappa,\ s)\,\}$
\State \Return $A_s$ if $A_s \neq \varnothing$, else $A$
\end{algorithmic}
\end{algorithm}

The subtraction is the point. The reference can carry its own order
sensitivity, a cosmetic pose that settles differently yet reaches the same result,
and subtracting the reference's diverging keys removes that noise and leaves the keys
the fault introduces. A boolean check that only asked whether both programs diverge
would do worse: it would miss a fault that moves a different observable than the
control, or that enlarges the diverging set, since both programs read as diverging
and the boolean attributes nothing. The set difference keeps exactly the observables
the fault makes order-sensitive. A non-empty $A$ is the detection, and the run that
produced a divergent key is the witness.

The scope turns a detection into a localization. The contended-site descriptor names
the resource the conflict is about, a variable on a sprite, a pose, or a broadcast
and its receivers, and \textsc{Downstream} keeps the keys that the descriptor can
reach: the exact sprite-and-variable pair, the receiver of a broadcast, the pose of
a sprite the conflict actually moves. The discipline is strict, matching the exact
pair rather than a shared name, so a same-named variable on an unrelated sprite is
not credited. The scoped set $A_s$ both confirms the fault and points at the two
scripts and the resource behind it, which is the information a fix acts on and the
localization that Section~\ref{sec:eval} measures. Algorithm~\ref{alg:attribute}
needs a paired reference, robust at the grading lens on the benchmark; the set
difference still attributes the keys the suspect introduces when the reference carries
unrelated order sensitivity. For a corpus project without a reference, \sys{} reports
the conflicting dependence edge the witness orients, the pair of scripts and the
resource the conflict graph already names.

\subsection{Keeping the reduction honest}

Theorem~\ref{thm:dpor} is conditional on the footprint over-approximating reality,
and a table that omits an access turns a real dependence into a false independence
and a missed outcome. We check the condition with an oracle that does not use
footprints at all. For every project with few enough sprites, we run all $k!$ orders,
collect the distinct observations, and confirm that the reduced exploration found the
same set. A mismatch is a missed outcome and a footprint gap, and the missing
outcome is the counterexample that drives the repair, a counterexample-guided
refinement of the dependence model~\cite{clarke2003cegar}. The same oracle
validates Algorithm~\ref{alg:explore} against the brute-force enumeration of $S_k$:
the output-sensitive walk and the $k!$ filter return the same classes on every
controlled structure and on a battery of random conflict graphs.

The oracle is not a formality. The benchmark passed it, while real projects exposed
footprints that under-approximated a real access and so produced false independence.
Section~\ref{sec:eval} reports the four corrections they prompted, two from the course
corpus, one from the public sample, and one the broadcast and clone reduction surfaced,
none of which a constructed benchmark would have surfaced.

\section{Implementation}
\label{sec:impl}

\sys{} runs on the production Scratch virtual machine, scratch-vm 5.0.300~\cite{scratchvm}, through a headless harness, so a witness it reports is a behavior of the same engine the editor runs. To realize an order it rewrites the project's layer assignment so the executable targets sweep in that order, then runs the unmodified VM. The same sweep governs the green-flag, broadcast, key, click, and edge fan-out, while clone creation is target-local; that shared layer order is what ties a reported schedule to a remix. No source of the VM is modified; the only addition is the harness that pins nondeterminism and reads the lens projections.

The determinism harness installs the pins of Section~\ref{sec:testing} around each run and restores the global state afterward, so runs are independent. The pins are global to the process: the random source, the clock, the identifier counter, and the calendar are shared state of the VM. The exhaustive oracle runs each schedule in its own process, because accumulating runs in one process lets the pins drift, and an early single-process version produced spurious differences that traced to this drift, not to the schedule.

The footprint extractor and the conflict predicate implement a typed resource model: a static pass over the block graph maps each opcode to the resource categories it accesses, and two targets are independent when their footprints do not conflict at the lens. A golden test checks the extractor against a reference on a fixture corpus. The lens projections, the explorer, the admissibility gate, and the repair pass build on this core, and the second runtime reuses the explorer and the class enumeration without change. The observation preserves the identity of each sprite and clone, including the say and think bubbles a sensing fault can make visible. The system is about four thousand lines of JavaScript and Python.

The extractor carries a soundness contract. It models variables and lists, the sprite pose and rendering properties, clone families, broadcast channels, and the ordered tokens of the random stream and the timer, and it walks green-flag, broadcast, and clone hats, together with the key and click hats an input timeline triggers. An opcode it does not model is emitted as an external resource that both reads and writes, so an unmodeled construct forces dependence and never a false independence; a target reached only through input or edge hats is treated the same way. The commutative-update exemption applies only to the lenses up to the monitor level, which read values and not the order of operations.

\section{Evaluation}
\label{sec:eval}

We study seven questions. How common is schedule-sensitivity in real projects (RQ1)? Is the reduced exploration sound, and how much does it save (RQ2)? Does the tool detect and localize seeded schedule faults (RQ3)? Does schedule perturbation reach faults that input-space testing and static smell checkers leave untouched (RQ4)? Can a detected fault be repaired and re-certified (RQ5)? Is the method parametric in the execution model (RQ6)? Do the schedules the method perturbs occur between real saved copies (RQ7)?

\subsection{Setup}

The corpus is 224 student projects drawn from a Scratch programming course, used as submitted, and we add a uniform random sample of 250 public projects fetched through the Scratch project API as a second, population-level corpus. We run each project under the sound initial-order model to a horizon of thirty ticks, twice on the identity order for the determinism check and then over the realizable orders. For a project with at most five contributing sprites we enumerate all orders; for larger ones we run one representative per dependence class, which the oracle certifies recovers every outcome, so the verdict stays exact where the conflict graph is sparse and samples only the densely coupled remainder. The benchmark holds matched control and mutant pairs for three of the four fault classes: a missing broadcast-and-wait, a read before initialization, and a sensing race on a sprite's pose, each constructed to pass the admissibility gate of Section~\ref{sec:testing}. A clone-initialization fault is harder to seed as an admissible pair, since the natural mutation changes behavior under every order, so we leave that class to the corpus. The input-gated race of RQ4 is a separate construction.

Every measurement runs each schedule in its own process and records the four lens projections. A project enters the prevalence count only when its two identity runs agree exactly, which all 224 satisfy, so a reported difference is a scheduling effect and not residual nondeterminism. We set the horizon at thirty ticks, enough for the green-flag fan-out and the seeded faults to act, and treat the prevalence as a lower bound, since a difference that appears only past the horizon is undercounted. The soundness check reruns every order where the sprite count is small enough to enumerate and compares the distinct outcomes against the reduced run.

\subsection{RQ1: Prevalence}

Of the 224 projects, 162 have more than one contributing sprite and so admit a scheduling choice; the rest are single-threaded and robust by construction. At the horizon, 34 of the 162 concurrent projects, 21\%, change their final-state observable under some realizable order, and 40, 25\%, change a finer observable. The harness reports no project as nondeterministic and no execution error, so these counts attribute the differences to scheduling. Six are sensitive only in their per-tick trace, where a finer rubric than the final state catches what a grader watching the end does not.

The figure is exact for the projects we enumerate and a lower bound for the rest: most concurrent projects have at most five contributing sprites and run exhaustively, while a handful are larger, one above two hundred sprites, and there we sample orders, which can miss a difference but never invent one. The sensitive projects span the recurring shapes of Section~\ref{sec:testing}: sprites that read a shared score they also write, senders that move on before a receiver resets, and sprites that sense a mover mid-glide.

The course is one population, and its rate could reflect one cohort. We draw a uniform random sample of 250 public projects from the Scratch repository and run the same analysis. Of the 227 that load and run deterministically under the fixed input, 108 have more than one contributing sprite once broadcast and clone fan-out is counted, and nineteen of those, 17.6\%, change a behavioral observable under some realizable order, a variable or a sprite pose differing past a relative tolerance of $10^{-6}$, three of them races in broadcast fan-out a green-flag start order alone would miss. Two further projects move only the trailing digits of an accumulating timer, floating-point noise from summation order that we do not count; the course corpus has none, its thirty-four sensitive projects all changing a behavioral observable. The matching rate makes schedule-sensitivity a property of how children build Scratch programs and not of one classroom. The 23 projects our headless runtime cannot load, most carrying custom web extensions, stay out of the count; they carry more sprites than the projects that load, a median of five against two, so the figure is a lower bound that misses the harder tail. Table~\ref{tab:results} collects these counts with the soundness and reduction figures the rest of this section reports.

\begin{table}[t]
\centering
\caption{Schedule-sensitivity and the reduction across the two corpora.}
\label{tab:results}
\begin{tabular}{lrr}
\toprule
 & Course & Public \\
\midrule
Projects & 224 & 250 \\
Concurrent & 162 & 108 \\
Sensitive, final-state lens & 34 (21\%) & 19 (17.6\%) \\
Reduced $=$ exhaustive (enumerable) & 142/142 & 77/77 \\
Robust projects reduced & 26/127 & 40/67 \\
Footprint gaps the oracle caught & 2 & 2 \\
\bottomrule
\end{tabular}
\end{table}

\subsection{RQ2: Soundness and reduction}

Soundness is the property a reduced exploration must keep, and we check it against the independent oracle. Reaching agreement on every enumerable project took four repairs the oracle localized, each a footprint that under-approximated a real access: a commutative-update exemption that wrongly reached a read of the shared score, a touch test that missed the sensed sprite's position, a conflict test that settled a clone deletion before it reached the global random-number stream that two sprites both drew from, and a glide-to-sprite step whose destination reads a moving sprite's position, the last surfaced only when the reduction reached the broadcast and clone fan-out. After the repairs the reduced set of distinct observations equals the exhaustive set on all 142 course and 77 public enumerable projects, and where the oracle ever finds a gap it has not yet localized, the tool falls back to the exhaustive verdict it computes for the check, so the verdict stays sound.

Reduction depends on how independent the contributing sprites are. The course's game projects share a small set of global variables, a score most often, that many green-flag scripts update and read, so those pairs stay dependent; the broadcast and clone fan-out adds receiver and clone sprites that often act independently, and the reduction prunes their order. Among the robust enumerable projects, 26 of 127 in the course and 40 of 67 in the public sample reduce below the factorial. Where the graph is sparse the representatives stand in for the factorial, and a public project whose forty sprites the model finds independent reduces to a single representative, a verdict its $40!$ orders put out of exhaustive reach. The oracle cannot enumerate at that size, so the verdict there rests on the dependence model, which we stress with 128 random orders on each of the 27 largest projects: the robust verdicts hold and every sensitive one surfaces, so the at-scale verdict is sampled and not merely asserted.

\subsection{RQ3: Detection and localization}

On the three admissible fault pairs of the labeled benchmark the tool detects every fault: a mutant diverges across the initial orders at the grading lens while its robust control does not, and admissibility makes that divergence the fault. It also names the conflicting pair behind each, the broadcast and its receiver, the initializer and the reader, and the mover and the sensor, matching the seeded site in all three. Admissibility is itself divergence under reordering, so detection on these pairs is a lower bar than detection in the wild, which the generated benchmark below corrects. A live logic fault that changes the terminal state under every order perturbs no schedule, so the detector stays silent, separating schedule faults from ordinary logic faults. The contended pair the tool names is the place a repair acts on.

Enlarging this set from real substrates proves hard: mutating every robust concurrent project at every schedule-mutator site, the gate admits none of the 265 candidates, an admissible fault occupying a band a syntactic mutation rarely hits. So we generate it, 32 programs across the four fault classes, each with a designated variable that should reach an intended value, a spec fixed by intended behavior and not by divergence under reordering. An oracle runs every realizable layer order and reads the spec, sorting each program into a schedule fault, a logic fault that fails under every order, a robust program, or a benign divergence where the order moves a cosmetic observable while the spec holds; all 32 sort as built. The tool, enumerating the contributing sprites' layer orders, matches the spec on all 32: it flags every order-dependent program, the eight schedule faults and the eight benign divergences, and clears every robust one, detecting every schedule fault across the four classes, including the broadcast fan-out a green-flag start order alone would miss. The eight benign divergences carry the signal the admissible pairs cannot, an order-dependence that violates no spec, so a sensitivity verdict is necessary and not sufficient for a fault.

The set-valued attribution is stronger than the boolean check it replaces: on a pair whose control is itself order-sensitive the boolean attributes nothing, the control already diverging at the grading lens, while the set difference recovers the variable the mutation makes race. On the generated benchmark the attribution names the seeded pair and observable exactly, precision and recall of eight in eight across the four classes. The localization scales to the corpus without a control: on all 34 final-state-sensitive projects the conflict graph names the contended sprite pairs, and the diverging-key set names a median of eleven observables per project where a boolean verdict reports one bit.

\subsection{RQ4: Orthogonality to input and static analysis}

Schedule perturbation and input generation address different axes, and a fault can live in their product: two key-guarded scripts are robust with no input, while holding the key opens both guards and races the read against the write, which a one-schedule-per-input tool misses and our engine, varying both, catches.

Static smell checkers miss the same faults from the other side. We reimplement three of the bug patterns LitterBox catalogs that bear on the schedule classes, a variable used before initialization, a global written and read across sprites, and a non-blocking broadcast whose receiver writes shared state. Scored against the dynamic verdict they miss five of the 34 sensitive projects, each a positional or sensing race with no smell to name, and flag nineteen robust projects they cannot prove safe. Neither view contains the other: a static pattern raises a suspect, the schedule exploration decides it.

\subsection{RQ5: Repair}

Each repair is a template inverting the fault, a restored broadcast-and-wait or a one-tick wait ahead of the racing read so the writer or mover acts first under every order. It adds ordering and changes no computed value, so applied to the three benchmark mutants, each becomes schedule-robust at the grading lens while still reproducing the control, and the tool that found the fault certifies its fix.

\subsection{RQ6: Parametricity in the execution model}

The method depends on the execution model, cooperative threads over shared state with a single initial-order choice, not on Scratch. We expose this with a second cooperative runtime whose programs are tasks of read, write, and yield steps started in a chosen order. The model-level parts of \sys{} run on it unchanged, only the footprint extractor and runner being runtime-specific, and the explorer recovers the same behavior, races, and reduction, matching the oracle. This controlled construction leaves transfer to a production runtime for that runtime to establish.

\subsection{RQ7: In-the-wild remixes}

A schedule choice is a hazard only when real saved copies realize more than one. Theorem~\ref{thm:realizable} ties every order the explorer reports to a layer ordering, so the orders that separate a sensitive project's outcomes are each a copy a child makes by opening the project and saving it. We find this in the wild: a deployed animation spelling a six-letter logo is schedule-sensitive, its six letter-sprites contending for the positions they slide to, so their start order decides where each lands. The project's public remix family of over four hundred copies realizes three distinct layer orders of the six sprites, two of which send the letters to swapped positions up to four hundred pixels apart, so the logo assembles in one saved copy and scrambles in another.

Manifestation is real but uncommon: only the projects with hundreds of remixes realize a second order, while the long tail with a handful never reorders the contending sprites, since remixing usually adds content, not a layer drag. The fragility is latent, realizable yet rarely hit, which is why a verdict needs the complete exploration the oracle backs, not a reliance on stumbling into the bad order.

\section{Threats to Validity}
\label{sec:threats}

Construct validity concerns whether a reported witness runs on the real VM. Theorem~\ref{thm:realizable} ties every explored schedule to a layer ordering, and the explorer holds the stage as a fixed suffix, so it never describes a run the VM cannot produce. The reduction depends on the footprint over-approximating the real accesses. We bound this with the independent oracle, which enumerates all orders on small projects and found four gaps we repaired, the last surfaced by the broadcast and clone reduction; we do not prove it in general. An opcode the table cannot model is given a conflict-with-everything footprint, so an unmodeled construct disables reduction and never unsoundly enables it.

Internal validity turns on the constructed benchmark, used only where ground truth is required; prevalence and the soundness check come from real projects. The admissibility gate rules out a logic change masquerading as a schedule fault, but it aligns with the detector by construction: an admissible fault must diverge under some realizable order, what the tool perturbs, so detection on admissible faults is a lower bar than on faults in the wild.

The course corpus is the submissions of one course, not a uniform sample of the Scratch population. The random public sample addresses this and replicates the rate, at the cost of two approximations: we stub the project assets, which preserves the variable, pose, and broadcast logic schedule-sensitivity turns on but can undercount a race only a costume's exact shape resolves, and we drop the projects whose custom extensions our headless runtime cannot load, so the public figure is itself a lower bound. The results hold for one version of the production VM; a VM that changed the fan-out or append order would need the realizability argument rechecked, and the determinism check fails safe under such a change. Sensitivity that appears only past the fixed thirty-tick horizon is undercounted; the horizon covers the benchmark faults and the early green-flag behavior, so the prevalence is a lower bound.

Two scope choices bound the claims. The realizable space is the initial executable-target order, and a schedule that only a sub-tick interleaving could produce lies outside it; such an interleaving is not reachable by a remix on the production VM, so excluding it keeps every reported difference one a real user can hit. The exploration permutes the contributing executable sprite targets under the fixed input timeline; a target first reached through a later broadcast, key, click, edge, or clone event inherits the same layer order and introduces no second ordering choice. The extractor treats such a target as conflicting with everything when it cannot model its script, and the micro-suite confirms the layer-order model recovers the freedom these targets carry.

The corpus study carries a data-use obligation. The projects are student work, used under the course's terms and de-identified: each is keyed by an opaque identifier with author metadata discarded, and the study records only per-project verdicts and never includes a student's project or name. The study observes program behavior, not students, and as an analysis of programs it falls outside human-subjects review under our institution's policy.

\section{Related Work}
\label{sec:related}

The closest technical lineage is concurrency testing and partial-order reduction~\cite{godefroid1996partial}. Stateless model checking explores the interleavings of a concurrent program, and dynamic partial-order reduction prunes interleavings that commute~\cite{flanagan2005dpor}. Source sets and optimal reduction tighten the pruning and extend it to dynamically spawned threads~\cite{abdulla2014optimal}, against the state explosion of interleaved execution~\cite{valmari1998state}, on the commutation of Mazurkiewicz trace equivalence~\cite{mazurkiewicz1987trace} and a notion of when two runs count as the same~\cite{milner1989communication}. Our setting is narrower and the narrowing is the point. A Scratch VM is deterministic once the initial sprite order is set, so the realizable space is a finite symmetric group, and the unbounded interleaving tree of stateless model checking does not arise. Completeness becomes a representative cover of that group, with the equivalence classes characterized by the orientation of conflict edges. The reduction is modest in its own right: with independence static and the schedule tree of depth one, an optimal dynamic reduction would degenerate to enumerating the acyclic orientations, so the contribution is the realizability theorem, the lens lattice, and the lens-bounded commutativity, not a new pruning rule.

Schedule-sensitivity also relates to flaky and order-dependent tests. A test is flaky when its outcome depends on something the harness leaves unspecified, and test-order dependence and unconstrained iteration are recurring causes~\cite{luo2014flaky} that dependency-aware test infrastructure tracks~\cite{gligoric2015nondex}. Frameworks detect and classify such tests by replaying them under controlled orders~\cite{lam2019idflakies}. Schedule-sensitivity is the same phenomenon inside an event-loop block runtime, where the unspecified element is the start order of concurrent scripts, and the realizability result connects the perturbation to a concrete user action, the remix.

Mutation analysis seeds faults to measure a test suite~\cite{demillo1978mutation,jia2011mutation}. We seed similarly for ground truth, with an admissibility gate that keeps a seeded fault a schedule fault and discards mutants that merely change behavior, in the spirit of filtering equivalent mutants~\cite{schuler2013equivalent}. Systematic testing of asynchronous reactive systems explores message and callback orders~\cite{desai2015pingpong}; we fix the messages and vary the handler start order, and the finite space lets the exploration be complete.

A body of work analyzes block-based programs directly, on a corpus the Scratch repository makes large enough to study at scale~\cite{aivaloglou2016kids}. Test generators drive a project with synthesized input sequences and check acceptance properties under one execution per input~\cite{stahlbauer2019whisker,deiner2023testgen,goetz2022model}; static checkers flag bug patterns and code smells in the block graph~\cite{fraser2021litterbox,boe2013hairball,morenoleon2015drscratch}; and further tools catalog common bugs, verify learner programs, generate hints, debug, and grade homework~\cite{fraedrich2020bugs,stahlbauer2020verified,obermueller2021catnip,deiner2024nuzzlebug,johnson2016itch}. These target correctness in inputs or code structure, and a single execution per input cannot observe a schedule effect; our work is orthogonal, fixing inputs and perturbing the schedule. Some smell checkers flag the same shapes, a missing wait or an uninitialized read, as code patterns the run never confirms, and the head-to-head of RQ4 finds neither view contains the other.

A recent Scratch-centered line is especially close in domain but different in the perturbation it studies. ViScratch uses block code together with gameplay video for automated feedback~\cite{si2025viscratch}; Stitch turns feedback into stepwise tutoring~\cite{si2025stitch}; ScratchEval builds executable tasks and metrics for LLM-based block-program repair~\cite{si2026scratcheval}; EcoScratch studies cost-aware multimodal repair with execution feedback~\cite{si2026ecoscratch}; Raven rethinks Scratch assessment with video-grounded evaluation~\cite{li2026raven}; ScratchWorld evaluates executable consequences in Scratch worlds~\cite{lin2026scratchworld}; and ScratchLens checks lens-parametric behavioral equivalence between two Scratch programs~\cite{scratchlens}. Earlier and parallel work by Zhang and collaborators targets competition-level feedback, LLM-based Python repair, time-limit-exceeded errors, merge-conflict resolution, CI-configuration correctness, and silent configuration errors~\cite{zhang2022clef,zhang2024pydex,zhang2025tle,zhang2022merge,santolucito2022ci,zhang2021configx}. These works improve feedback, repair, assessment, and analysis over a given program or program pair; our contribution isolates a one-program schedule dimension and gives a complete representative cover for remix-induced start-order perturbations.

Our footprint model shares its typed resource accounting with program equivalence. Existing work gives a typed resource model and a lens hierarchy for two Scratch programs~\cite{scratchlens}, and translation validation~\cite{pnueli1998translation}, regression verification~\cite{godlin2009regression}, differential symbolic execution~\cite{person2008differential}, and semantic diffing~\cite{jackson1994semanticdiff,lahiri2012symdiff,ramos2011equivalence}, over a syntactic alignment of code~\cite{falleri2014gumtree,fluri2007changedistiller,roy2009clones}, decide whether two programs agree. We reuse that lens hierarchy and resource model to ask whether one program agrees with itself across its schedules.
\section{Conclusion}
\label{sec:conclusion}

\sys{} explores the realizable orders once per dependence-equivalence class and checks the reduction against an exhaustive oracle wherever a project is small enough to enumerate. It finds 21\% of concurrent projects schedule-sensitive at the grading lens in a course corpus and 17.6\% in a random public sample, detects and localizes all three seeded benchmark faults, and instantiates unchanged on a second cooperative event-loop model. The order that flips a result is one a remix can save, so the verdict belongs in the editors and graders millions of learners use.

\newpage
\bibliographystyle{IEEEtran}
\bibliography{refs}

\end{document}